\documentclass[a4paper,twocolumn]{esapub}
\usepackage{times,natbib,psfig}

\newcommand{\SSS}{\renewcommand{\baselinestretch}{1.0}\tiny \normalsize}
\newcommand{\SSH}{\renewcommand{\baselinestretch}{1.2}\tiny \normalsize}

\newcommand{\kms}{\mbox{km\,s$^{-1}$}}

\newcommand{\pccm}{\mbox{cm$^{3}$}}
\newcommand{\rj}{\mbox{R$_{\rm J}$}}
\newcommand{\Lower}[1]{\smash{\lower 1.5ex \hbox{#1}}}

\def\chem{\everymath={\fam0 }\fam0 }

\def\underlinewords#1{%
	\def\stuff{#1 }\leavevmode\expandafter\ulword\stuff * }
\def\ulword#1 {\def\one{#1} \ifx\one\aster\let\next\relax
\else\vtop{\hbox{\strut#1}\hrule\relax}
\let\next\ulword\fi\next}
\def\strikeoutwords#1{%
\def\stuff{#1 }\leavvmode\expandafter\soword\stuff * }
\def\soword#1 {\def\one{#1} \ifx\one\aster\let\next\relax
\else\vtop{\hbox{\strut#1}\kern-.5\baselineskip\hrule\relax}
\let\next\soword\fi\next}
\def\aster{*}

\begin{document}


\title{Io Revealed in the Jovian Dust Streams}

\author{Amara Graps}
\author{Eberhard Gr\"un}
\author{Harald Kr\"uger}
\affil{Max-Planck-Institut f\"ur Kernphysik, Heidelberg, Germany}
\author{Mih\'{a}{\l}y Hor\'anyi}
\affil{LASP, Boulder, USA}
\author{H{\aa}kan Svedhem}
\affil{ESTEC, Noordwijk, Netherlands}


%

\maketitle


\begin{abstract}

The Jovian dust streams are high-speed bursts of submicron-sized
particles traveling in the same direction from a source in the Jovian
system.  Since their discovery in 1992, they have been observed by
three spacecraft: Ulysses, Galileo and Cassini.  The source of the
Jovian dust streams is dust from Io's volcanoes.  The charged and
traveling dust stream particles have particular signatures in
frequency space and in real space. The frequency-transformed Galileo
dust stream measurements show different signatures, varying
orbit-to-orbit during Galileo's first 29 orbits around Jupiter.
Time-frequency analysis demonstrates that Io is a localized source of
charged dust particles. Aspects of the particles' dynamics can be seen
in the December~2000 joint Galileo-Cassini dust stream measurements.
To match the travel times, the smallest dust particles could have the
following range of parameters: radius: 6~nm, density:
1.35--1.75~g/\pccm, sulfur charging conditions, which produce dust
stream speeds: 220$\backslash$450~\kms\ (Galileo$\backslash$Cassini)
and charge potentials: 5.5$\backslash$6.3~V
(Galileo$\backslash$Cassini).

\end{abstract}

\section{Overview}

The Jovian dust streams are high-speed collimated streams of
submicron-sized particles traveling in the same direction from a
source in the Jovian system.  They were discovered in March~1992 by
the cosmic dust detector instrument onboard the Ulysses spacecraft,
when the spacecraft was just past its closest approach to Jupiter.
Observations of the Jovian dust stream phenomena continued in the next
nine years. A second spacecraft, Galileo, now in orbit around Jupiter,
is equipped with an identical dust detector instrument to Ulysses'
dust instrument.  Before and since the Galileo spacecraft's arrival in
the Jupiter system in December~1995, investigators recorded more dust
stream observations. In July and August 2000, a third spacecraft with
a dust detector (combined with a chemical analyzer), Cassini,
traveling on its way to Saturn, recorded more high-speed streams of
submicron-sized particles from the Jovian system. The many years-long
successful Jovian dust streams observations reached a pinnacle on
December~30,~2000, when both the Cassini and Galileo dust detectors
accomplished a coordinated set of measurements of the Jovian dust
streams inside and outside of Jupiter's magnetosphere.

Indirect methods applied by previous researchers have pointed to Io
being the simplest explanation for the question of the origin of the
Jovian dust streams. We first show by \underline{direct} methods that Io is the
source of the Jovian dust streams. To address the issue of identifying
Io directly in the Galileo dust detector data, we apply time-frequency
analysis, in particular, Fourier methods, to the Galileo dust data.
Additional frequency signatures, such as amplitude modulation, also
emerge from the time-frequency analysis.

The second part of this paper focuses on the dust streams dynamics.
Here, we apply a detailed Jovian particles and fields model to
simulate a dust stream particle's trajectory as the particle moves
from Io's orbit through Jupiter's magnetosphere and beyond. Through
the model, we show one possible set of parameters that match the
travel times seen in the Dec\-ember~30,~2000 Galileo-Cassini joint
dust stream measurements.

\section{Io's Frequency Fingerprint}

In order to find Io's frequency `fingerprint' in the Galileo dust
detector data, we followed the following steps: 1)~we transformed the
Galileo dust detector data into frequency space via periodograms,
2)~we noted frequency patterns such as amplitude modulations, 3)~we
compared the frequency-transformed Galileo data with synthetic data,
and, 4)~we noted spacecraft effects such as Doppler shifts.

\subsection{Time-Frequency Analysis via Periodograms}

The ``classic" or Schuster periodogram \citep{Brett:88}\ is
conventionally defined as the modulus-squared of the discrete Fourier
transform.  If the input time series contains a periodic feature, then
the periodogram can be calculated for any frequency and it displays
the presence of a sinusoid near one frequency value as a distinct peak
in the spectrum. The Lomb-Scargle periodogram applied here is a
slightly modified version of the classic periodogram giving a simpler
statistical behavior \citep{Scargle:82}.

\begin{figure}[htb]
\psnodraftbox
\psfig{figure=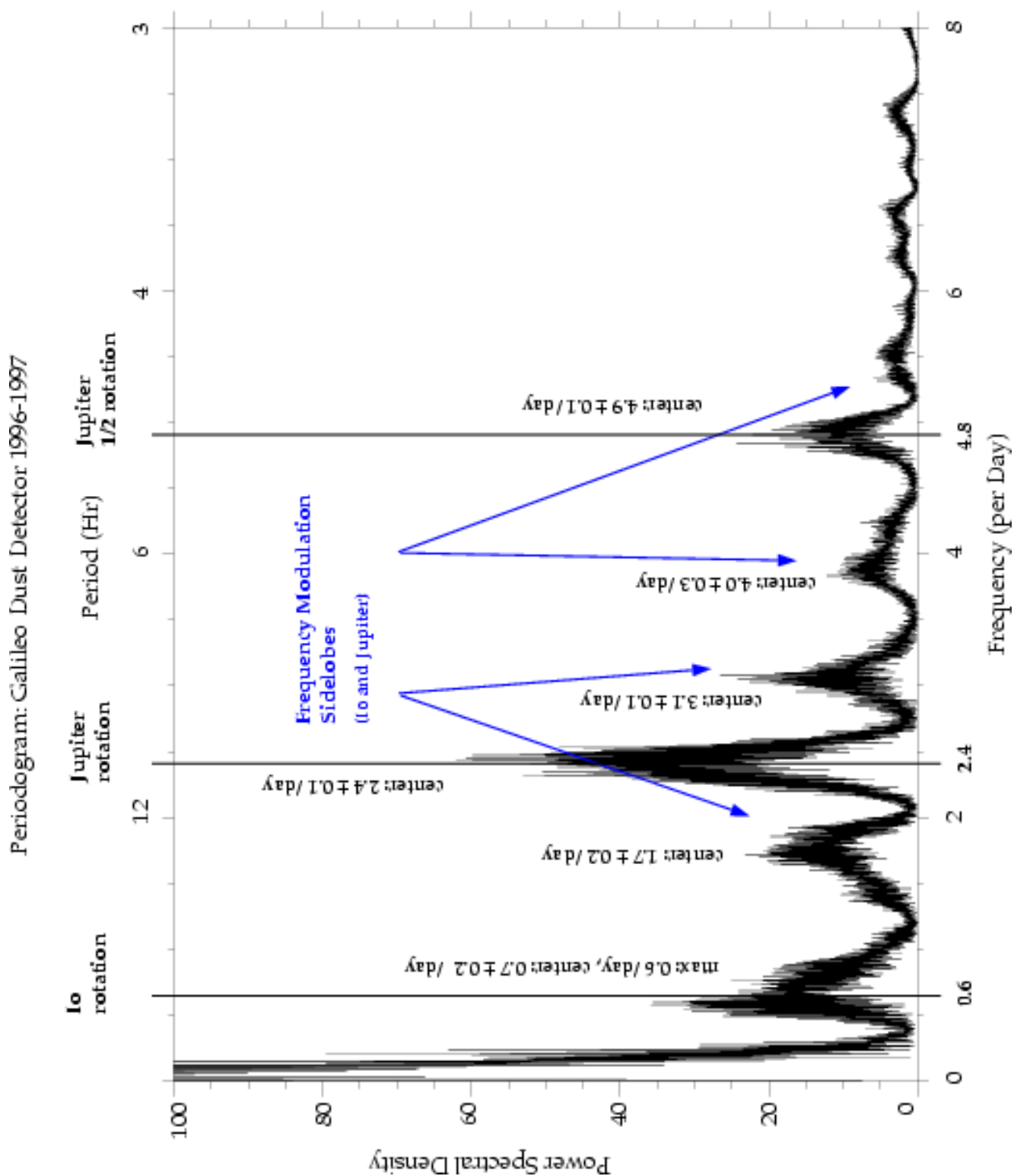,height=8.0in}
\SSS
\caption[Lomb-Scargle periodogram of Galileo 1996-1997 rate data,
noting frequency signatures.] {A Lomb-Scargle periodogram for the
first two years, 1996-1997, of Galileo dust impact rate data, with the
specific signatures described in the text. Io's frequency of orbital
rotation, in particular, can be seen here in frequency space, which
provides our first direct evidence that Io is a source of the Jovian
dust streams. A confirmation of Io's role as a localized charged dust
source arises through the modulation effects. The vertical solid lines
mark Io's orbital and Jupiter's rotational periods, and the arrows
point to Jupiter's modulation products with Io straddling Jupiter's
frequency.  The first harmonic of Jupiter's rotation frequency is
visible at ($\omega_1$ = 4.8~day$^{-1}$) and Jupiter's modulation
products with Io, which are straddling that first harmonic peak, can
be seen, as well. The strong frequency peak near the origin at 1 over
Galileo's orbital period is due to the Galileo spacecraft orbital
geometry.}
\label{periodogram}
\SSH
\end{figure}

Our best Galileo dust dataset for detecting Io's frequency fingerprint
emerged from the earlier Galileo orbits around Jupiter, because the
spacecraft orbital geometry of the first years of the Galileo mission
favored higher fluxes of dust stream particles. Therefore, by
combining two years of the early data, we gained a higher
signal-to-noise dataset. In Fig.~\ref{periodogram},
we show a Lomb-Scargle periodogram for the first two years, 1996-1997,
of Galileo dust impact rate data.  This particular periodogram is the
best example from the Galileo dust detector data showing, with high
confidence: Io's frequency signature, Jupiter's frequency signatures,
and amplitude modulation effects. The
periodogram shows the following frequency signatures.

\smallskip
\noindent{\underline{Frequency Summary of Fig.~\ref{periodogram}}}

\begin{enumerate}
\item A strong peak near the origin,
\item An asymmetric peak:
maximum at 0.6~day$^{-1}$, center at 0.7~$\pm$~0.2~day$^{-1}$,
\item An
asymmetric peak: center at 1.7~$\pm$~0.2~day$^{-1}$,
\item A tall peak:
center at 2.4~$\pm$~0.1~day$^{-1}$,
\item A peak: center at
3.1~$\pm$~0.1~day$^{-1}$,
\item Harmonics of the previous three peaks,
and
\item Progressively smaller and less-defined peaks.
\end{enumerate}

\subsection{Amplitude Modulation in Frequency Space}

Frequencies in frequency space can interact in numerous ways. We
interpret the frequency peaks, seen in Fig.~\ref{periodogram}, to be
the result of Io's frequency of orbital rotation, Jupiter's magnetic
field frequency of rotation, and an interaction between these two
frequencies called amplitude modulation (AM). The simplest case of AM
is a sinusoid modulating the amplitude of a carrier signal, which is
itself a sinusoid. Then the carrier signal is broken down in frequency
space into several sinusoidal oscillations: $x\approx \sin (\omega
_0t)+\sin (\omega _0t)\sin (\Omega t)$, which can be converted to sums
of frequencies using a trigonometric identity for sine products. The
result is a signature in frequency space that displays a carrier
frequency: $\omega_0$ with side frequencies (``modulation products''):
($\omega_0+\Omega$) and ($\omega_0-\Omega$).

The process of amplitude modulation applies to Io's orbital and
Jupiter's rotational frequencies in the following way. Jupiter's
rotation period is 9.8~hours corresponding to a frequency of
2.4~rotations per day. Io's orbital period (and rotation period) is
1.8~days, corresponding to a frequency of 0.6~rotations per day. If
the dust originates from Io, and the dust flux is modulated by
Jupiter, then the spectrum in frequency space would appear like the
spectrum in Fig.~\ref{periodogram}, where the modulation products
(sidelobes) at Jupiter's frequency at full and half-rotations are due
to Jupiter's frequency modulating Io's frequency of orbital
rotation. The frequency difference between Jupiter's rotational
frequency and each of the sidelobes is the same frequency as Io's
frequency of orbital rotation.  In addition, if one or both of the
original signals have broad spectra, then the spectrum of the
modulation products will be broadened to the same extent (since the
modulation process is linear).  Therefore, the spread of Io's peak is
repeated in the same way for the sidelobes of Jupiter's frequency
peaks (full and half-rotation).

Modulation products appeared in about 20\% of the first 29
frequency-transformed Galileo orbits. Galileo's orbit~E4 was the first
orbit for which a hint of the amplitude modulation appeared, then in
orbit~G8, the signature was unmistakable. Other Galileo spacecraft
orbits for which one can clearly see the modulation products are: C10,
E18, and G29.  Periodograms of each of the individual orbits can be
seen in \citet{Graps:2001}.

\subsection{Synthesizing the Frequency-transformed Dust Stream Data}

\begin{figure}[htb]
\psnodraftbox
\hspace*{.5in}
\psfig{figure=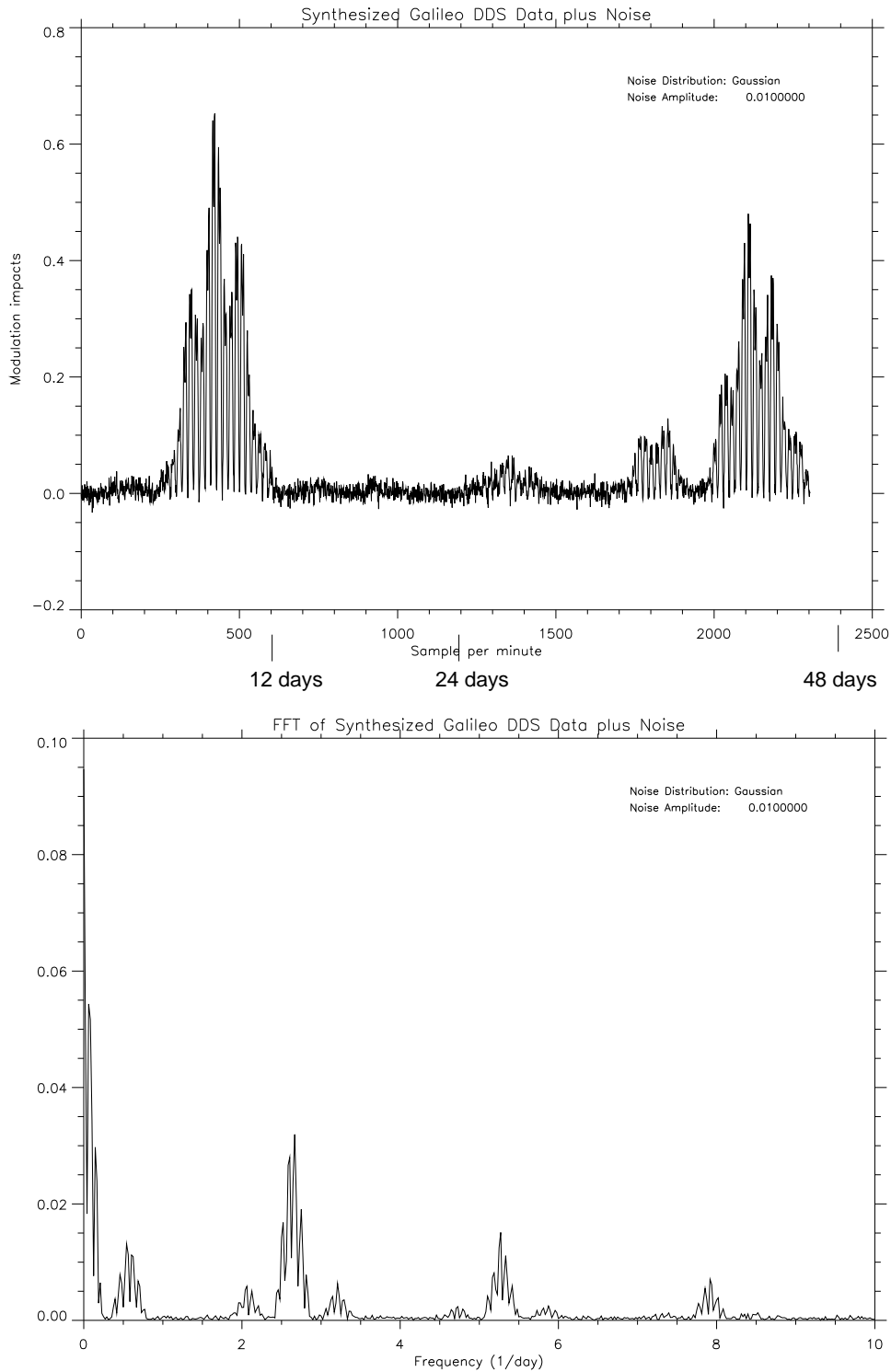,width=5.0in}
\SSS
\caption[A synthetic dataset and frequency-transformed.]
{Top: A synthetic impact rate dataset, with added Gaussian noise,
which includes the following periods: 48~days for Galileo's orbit, 9
and 17~days for smaller data blocks, a Jupiter rotation and half
rotation period, and an Io rotation period.  Bottom: An FFT of the synthetic
rate dataset.  Galileo's motion (at the origin),
Io's frequency of orbital rotation (0.6~d$^{-1}$), Jupiter's full
frequency (2.4~d$^{-1}$) and half-frequency (4.8~d$^{-1}$) of rotation
can be seen, as well as the Io-Jupiter amplitude modulation products
(sidelobes around Jupiter's frequencies).  }
\label{synthetic}
\SSH
\end{figure}

The main physical processes behind the frequency-transformed data can
become clearer when one synthesizes data and compares the synthesized
data to the real data. We have synthesized a spectrum, using typical
periods from Galileo's dust impact rate data, a Jupiter rotation and
half rotation period, and an Io orbital rotational period, which show
in the real data.  We added some Gaussian-distributed noise, and after
transforming the synthesized data into frequency space with an FFT,
frequency peaks appear with their modulation products, which are in
the same locations as those in the 1996-1997 Galileo dust detector
periodogram. Figures of the synthesized time series and
frequency-transformed data can be seen in Fig.~\ref{synthetic}.

\subsection{Spacecraft Effects in Frequency Space}

Several Galileo spacecraft orbital characteristics can be identified
in the frequency-transformed data. The first effect is at the origin.
For each orbit, the dust instrument receives more dust impacts while
in the inner Jovian system than while in the outer Jovian system,
which in frequency space, results in a peak at the origin. The second
spacecraft effect is a Doppler effect between Galileo and Io, as the
``observer" and the ``source." In frequency space, the result is that
both the Io and the Jupiter peaks can be smeared by Doppler shifts.
The Io frequency peak Doppler shift is to shorter periods. This
asymmetry appears in the modulation products, as well. A table of the
Doppler shift trends can be seen in \citet{Graps:2001}.

\subsection{August-September 2000 Dust Storm}

\begin{figure}[htb]
\psnodraftbox
\hspace*{.5in}
\psfig{figure=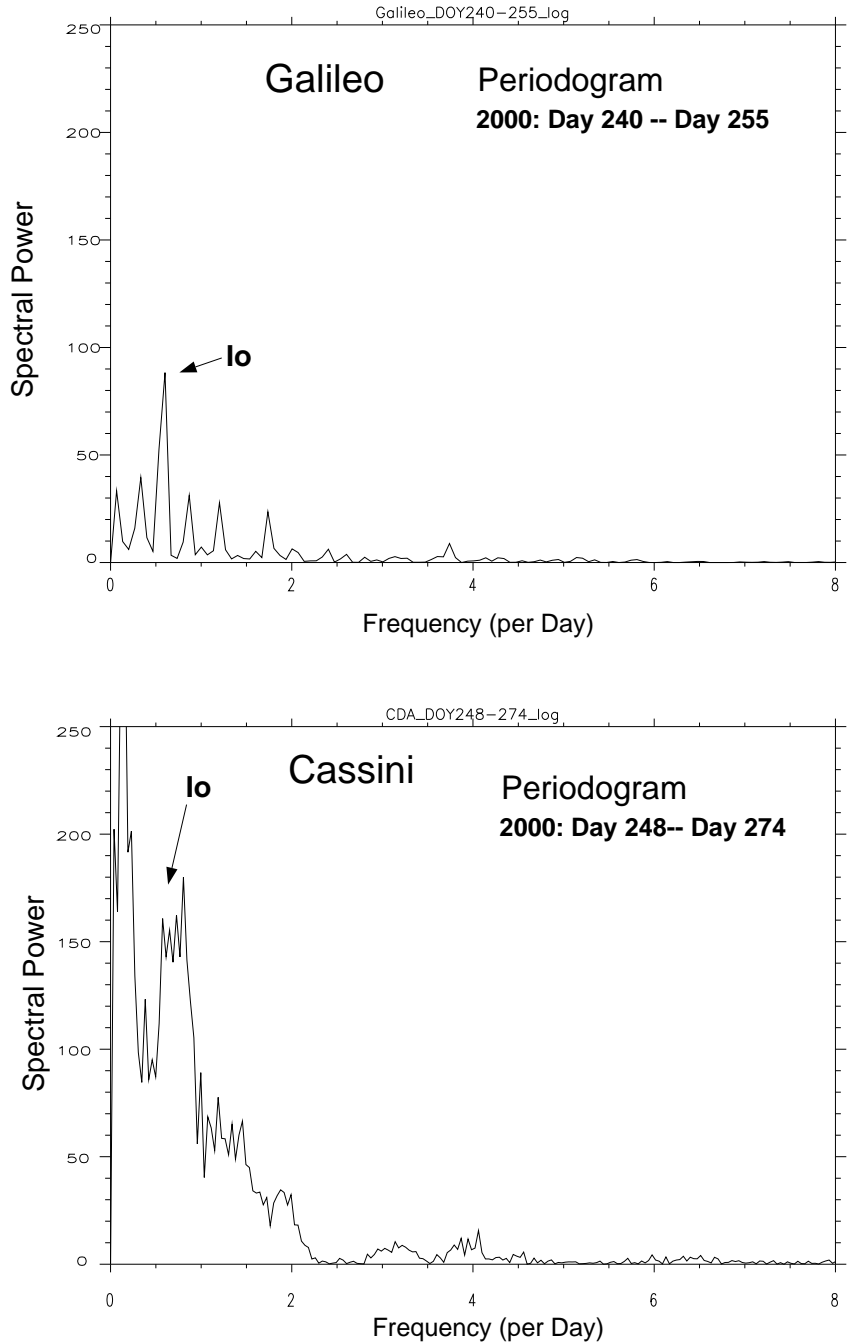,height=7.0in}
\SSS
\caption[Lomb-Scargle periodogram of Summer 2000 dust storm] {A
Lomb-Scargle periodogram of dust impact rate data from Galileo (top)
and Cassini (bottom) during selected days in July and August 2000. The
Galileo data occurred during Galileo's G28 orbit, when the spacecraft
was located outside of Jupiter's magnetosphere at $\sim$250~\rj (where
\rj=7.134~10$^{9}$~cm is the equatorial radius of Jupiter). This Io
dust streams storm detected in interplanetary space was stronger in
the earlier portion of this orbit close to perijove on Days~218--240
than in the later days, when Cassini detected the storm. Cassini
observed high impact rates particularly on $\sim$Day~251 and
$\sim$Day~266, when the spacecraft was traveling far outside of
Jupiter's magnetosphere at $\sim$1~A.U.}
\label{storm}
\SSH
\end{figure}

In August and September~2000, both Cassini (travelling by at
$\sim$1~A.U. from Jupiter), and Galileo (in an orbit carried to
$\sim$250~\rj\ from Jupiter, where \rj=7.134~10$^{9}$~cm is the
equatorial radius of Jupiter) detected increases in the rate of dust
impacts, which were approximately 100~times their nominal impact
rates. In the frequency-transformed data from both spacecraft, Io's
frequency signature swamped all other frequency signatures in the
Galileo data, which was noteworthy, because both spacecraft were
located far from Jupiter, outside of Jupiter's magnetosphere.  Galileo
detected the `dust storm' earlier during the perijove portion of its
G28~orbit, during Days~$\sim$218--240. After Day~240, Galileo's impact
rate decreased, however, Cassini observed high impact rates
particularly on $\sim$Day~251 and $\sim$Day~266. (See the periodogram
of the dust impact rate in Fig.~\ref{storm}). Cassini's dust
detector's observational geometry is very different from Galileo's,
however, by accident, Cassini captured the dust storm approximately
1-2 weeks after Galileo detected the dust storm.

\subsection{Conclusions of Io's Frequency Fingerprint}

Frequency analysis via Fourier techniques of the Galileo dust data
provides our first direct evidence of an Io dust source.  The presence
of Io's rotation frequency argues that Io is a localized source of
charged dust particles because charged dust from diffuse sources would
couple to Jupiter's magnetic field and appear in frequency space with
Jupiter's rotation frequency and its harmonics. A confirmation of Io's
role as a localized charged dust source arises through the modulation
effects.

The Galileo dust detector periodogram data shows variabilities,
orbit-to-orbit, even for some orbits which share similar orbital
geometry. This orbit-to-orbit variability is a clue that, either the
intervening medium or Io itself, i.e. its volcanoes, is a source of
the variability.

\section{Jovian Dust Stream Dynamics}

To describe the dynamics of the dust streams, we apply a detailed
Jovian particles and fields model to simulate a dust stream particle's
trajectory as the particle moves from Io's orbit through Jupiter's
magnetosphere and beyond. For the model, one needs to assume
approximations for the following: 1)~Jupiter's magnetic field,
2)~Jupiter's plasma, 3)~dust particle density, 4)~dust particle
optical property, 5)~charging processes, and 6)~forces.

\subsection{Jovian Dust Stream Model Details}

We approximate Jupiter's magnetic field by implementing Connerney's
$\chem O_4$ or $\chem O_6$ model \citep{Conn:81a,Conn:93}, which is a
quadrupole expansion of the planet's internal field. Additionally, we
hinge Connerney's current sheet described in \citet{Conn:81b}, to approximate
Jupiter's magnetodisk. Connerney's sheet implementation considers the
magnetodisk as a perturbation to Jupiter's internal field.

Jupiter's plasma is approximated using a plasma model which is a fit
to the Voyager~1 and 2 cold plasma measurements, described in
\citet{Bag:89}. We assumed a constant mixing ratio of 50\% between
single ionized oxygen and sulfur ions.

For the dust particle density, we use density values
1.35--2.0~g/\pccm. The dust particle's optical properties are
manifested via $Q_{pr}$,  which affects the particle's dynamics
through the radiation pressure force. For this work, a $Q_{pr}$ value
for the dust particle is calculated based on the particle's size, and
following the curve in Burns et al.'s classic paper: \citet{Burns:79}.

The charge of the dust particles is approximated by summing over the
currents: photoelectron emission, ion and electron collection, and
secondary electron emission. The charge of the particle, which varies
in time, is integrated simultaneously with the particle's
acceleration. Here, we model the acceleration by considering the
following forces: Jupiter's gravitational force, the light pressure
force, the Lorentz force, and the solar gravitational force
\citep{Hor:97}. We have neglected the neutral gas and plasma (Coulomb)
drag forces on the dust particle because the time-scales used in these
runs are short (hours to a day), compared to the time-scales over
which those forces have an effect.

More details of the Jovian dust stream model  can be found in
M.~Hor\'anyi's papers, such as \citet{Hor:97}, and in
\citet{Graps:2001}.

\subsection{Dynamics Results}

A window of particle sizes exists for which dust particles can
escape from traveling in Keplerian orbits in Jupiter's magnetosphere.
For small dust particles, their motion is as plasma ions and
electrons, which gyrate about Jupiter's magnetic field lines. For
large dust particles, their motion is governed by gravity. From Io's
orbital location, the window of dust particle sizes for escaping
particles is approximately radius 5~nm to 35~nm. This particle size is
strongly dependent on the charging assumptions, especially the
secondary electron emission material assumption. In our numerical
experiments thus far, we found that smaller dust particles can be
ejected when the impacting ion or electron energy is lower than the
energy of other impacting ions or electrons in the secondary electron
emission process. For example, a particle with sulfur properties can
be ejected from Io's location with a minimum size 4~nm, versus a particle
with silicate material properties, which can be ejected from Io's location
with a minimum size 6~nm.

As the dust stream particle moves through the Jovian magnetosphere,
equilibrium potential is rarely reached. Therefore, as the particle
moves outward, it continues to collect charges, which further
accelerate the particle. We now think that the Jovian dust stream
particles move faster (at least 400~km/sec) than previously
assumed in the earlier work presented by \citet{Zook:96}, which
suggested dust speeds, at least 200~km/sec.

\subsection{Joint Measurements: December~2000 from Galileo \& Cassini}

\begin{figure}[h]
\psnodraftbox
\hspace*{.5in}
\psfig{figure=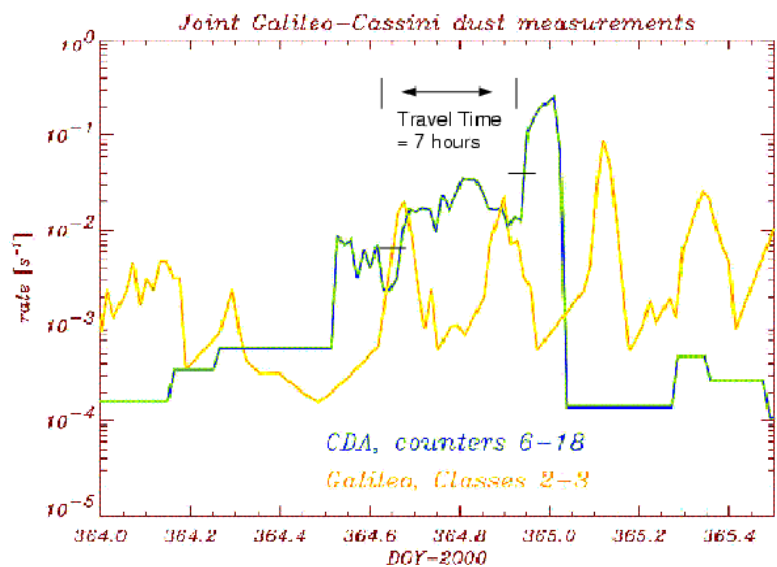,width=4.0in}
\SSS
\caption[Dust impact rates from the Cassini-Galileo dual dust stream
measurements performed on December 30,~2000.]{Dust impact rates from
the Cassini-Galileo dual dust stream measurements performed on
December 30,~2000. The marked travel time, 7~hours, is the time for a
dust particle in a collimated dust stream to travel from Galileo to
Cassini.}
\label{joint}
\SSH
\end{figure}

\begin{figure}[htb]
\psnodraftbox
\hspace*{.5in}
\psfig{figure=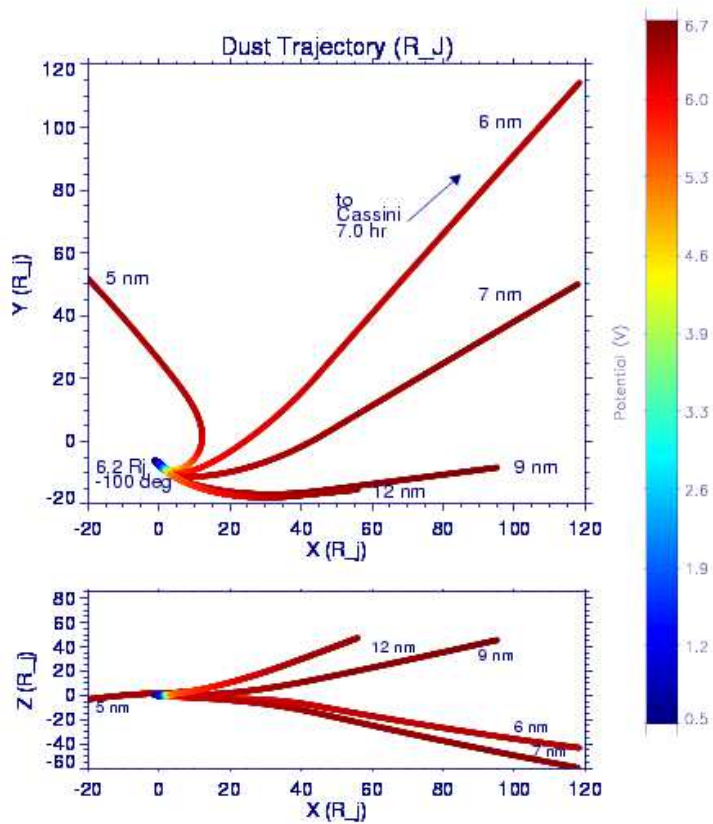,width=4.0in}
\SSS
\caption[Dust particle trajectories for different sized particles released
at 6.2~Jovian radii.]{Dust particle trajectories in x-y (top) and x-z (bottom)
Jupiter-centered coordinates for different sized particles released
from a near-Io Keplerian orbit at 6.2~Jovian radii. The colors indicate the
values of the particle's charge potentials. The trajectory with the
arrow indicates a reasonable trajectory which would bring the
particle near to the Cassini spacecraft.}
\label{phase100}
\SSH
\end{figure}

On December 30,~2000, the Cassini spacecraft closely flew by Jupiter,
providing a simultaneous two-spacecraft measurement (Cassini-Galileo)
of particles from one collimated stream from the Jovian dust streams.
Particles in a stream were detected by Galileo, as the spacecraft was
orbiting inside of the Jovian magnetosphere close to Ganymede
(8--12~Jovian radii), and then particles in the stream traveled to Cassini, as
Cassini flew by Jupiter at approximately 140~Jovian radii. Figure~\ref{joint}
shows the dust impact rate data for the dual dust stream measurements,
the gold line denotes the Galileo rates and the green line indicates
the Cassini rates. We assumed that the same dust stream at each
spacecraft began where the black horizontal line marks the midpoint of
the peak rise in impact rate. The travel time between the two
black-marked peaks is approximately 7~hours. One goal of the
dynamical modeling was to match this travel time.
From preliminary modeling, Fig.~\ref{phase100} shows the result of one possible
trajectory of a Jovian dust streams particle released near Io's orbit.
Here, the smallest dust particles could have the
following range of parameters: size: 6~nanometers, density:
1.35--1.75~g/\pccm, initial charge potential: 1--4~V, secondary
electron emission yield: 3.0, dependent on a maximum electron energy
300~eV, and a photoelectron emission yield: 0.1--1.0, which
produce dust particle speeds: 220$\backslash$450~\kms\
(Galileo$\backslash$Cassini) and charge
potentials: 5.5$\backslash$6.3~V (Galileo$\backslash$Cassini).
Smaller and larger particles than 6~nanometers result in the wrong direction towards
Cassini, and with travel times that are either too fast or too slow.

\section{Synopsis}

Our work from frequency analysis shows Io as the dominant source of
the Jovian dust streams.  The
variability seen in the frequency analysis shows that we might be able
to use dust stream measurements to monitor Io's volcanoes' plume
activity.

Our charging and dynamics modeling, (more details in
\citet{Graps:2001}), shows that the dust streams' equilibrium potential
is rarely reached in the Jovian magnetosphere, and that the Jovian
dust stream particles travel faster than found previously in
\citet{Zook:96}. In our preliminary analysis and modeling of the
Galileo-Cassini dust stream measurements, we show that one set of
conditions, which can match the travel times, gives a dust streams
speed at Galileo of 220~\kms, with a charge potential: 5.5~V, and a
dust streams speed at Cassini of 450~\kms, with a charge potential:
6.3~V.

\section*{Acknowledgements}
The authors gratefully acknowledge the hard work of the Galileo and
Cassini Dust Science Teams. Funding provided by the Deutsches Zentrum
f\"ur Luft-und Raumfahrt E.V. (DLR), and the Deutsche
Forschungsgemeinschaft (DFG).

\bibliographystyle{aa}
\bibliography{amthesis}

\vfil\eject

\onecolumn

\end{document}